\documentclass[%
 floatfix,
 preprint,
 superscriptaddress,
 amsmath,amssymb,
 pra,
]{revtex4-1}

\usepackage{graphicx}
\usepackage{dcolumn}
\usepackage{bm}
\usepackage{longtable}

\begin{document}
\title{Critical evaluation of Breit and QED effects on the $3d^9\ ^2D_{3/2}\rightarrow ^2D_{5/2}$ transition energy in Co-like ions.}

\author{R. Si}
 \affiliation{Department of Computer Science, University of British Columbia, Vancouver, Canada V6T 1Z4}
\author{X. L. Guo}
\affiliation{Department of Radiotherapy, Changhai Hospital, Second Military Medical University, Shanghai, China 200433}
\author{T. Brage}\email{Tomas.Brage@fysik.lu.se}
\affiliation{Division of Mathmatical Physics, Department of Physics, Lund University, 221 00 Lund, Sweden}
\affiliation{Shanghai EBIT Laboratory, Key Laboratory of Nuclear Physics and Ion-beam Application, Institute of Modern Physics, Department of Nuclear Science and Technology, Fudan University, Shanghai, China 200433}
\author{C. Y. Chen}\email{chychen@fudan.edu.cn}
\affiliation{Shanghai EBIT Laboratory, Key Laboratory of Nuclear Physics and Ion-beam Application, Institute of Modern Physics, Department of Nuclear Science and Technology, Fudan University, Shanghai, China 200433}
\author{R. Hutton}\email{rhutton@fudan.edu.cn}
\affiliation{Shanghai EBIT Laboratory, Key Laboratory of Nuclear Physics and Ion-beam Application, Institute of Modern Physics, Department of Nuclear Science and Technology, Fudan University, Shanghai, China 200433}
\author{C. Froese Fischer}
 \affiliation{Department of Computer Science, University of British Columbia, Vancouver, Canada V6T 1Z4}

\date{\today}

\begin{abstract}
The multiconfiguration Dirac-Hartree-Fock theory is used to calculate the $3d^9\ ^2D_{3/2}\rightarrow ^2D_{5/2}$ transition energy for Co-like ions with $Z=28-100$.
We investigate how electron correlation, frequency-independent and -dependent Breit interaction, as well as QED corrections vary along the sequence.
The well understood frequency-independent Breit contribution has the largest contribution for all ions. Among the corrections to this, correlation decreases rapidly with $Z$, the frequency-dependent Breit contribution is important especially for high-$Z$ ions, and the Self-energy contribution to the QED becomes the largest correction already for $Z>50$.
We evaluate and compare results for the Self-energy in three different approximations, (i) the approach implemented in the GRASP2K package, (ii) the method based on Welton's concept and (iii) a model operator approach recently developed by Shabaev and co-workers.
Through comparison with experimental values, it seems that the third set of results have the best agreement with experiments, but the difference from experiments for high-$Z$ ions, is around 0.03\%-0.04\%, and therefore our results are outside the error bars of the experiments.
\end{abstract}

\maketitle

\section{\label{sec:introduction}Introduction}
In calculation of accurate transition energies for atoms and ions, we are faced with essentially two challenges. First the contribution from electron correlation and, second, the Breit and Quantum-Electrodynamic (QED) effects. If we therefore would like to study the latter accurately in many-electron systems, it is important to choose systems where correlation contributions are small. Recently some of the authors proposed the ground configurations of F-like ($2p^5$) ions for this purpose~\cite{Li2018}, being what we labeled as a "Layzer-quenched" case. This implies that
we predict small contributions from correlation, since the $2p^5\ ^2P$ represent the only term in its Layzer-complex~\cite{Layzer1959,Layzer1962}. In this paper we report on a similar investigation for another Layzer-quenched system - Co-like with a $3d^9\ ^2D$ ground term.

Co-like systems were recently investigated in an elaborate treatment of correlation using the multiconfiguration Dirac-Hartree-Fock (MCDHF) method~\cite{Fischer2016} with the GRASP2K package~\cite{Jonsson2013}, giving fine-structure energy splittings of the ground term for $28\leq Z\leq100$~\cite{Guo2016}. However, in spite of showing excellent agreement with experiment, these calculations were limited for two reasons - first, the Breit interaction was included using the low-frequency limit for the exchanged photon frequency $\omega_{ij}\rightarrow0$,
discarding the frequency-dependent part
(see Theory-section below). Secondly, the Self-energy correction (the dominating part of the QED effect for these ions) was only included via the standard approach in GRASP2K, using the hydrogenic results of Mohr et al.~\cite{Mohr1983} and Klarsfeld et al.~\cite{Klarsfeld1973}.
In this, the screening effect is included through a screened nuclear charge by taking the overlap integral of the GRASP2K-wavefunction and a hydrogenic wavefunction.
It is clear that this approach opens up for improvements, both regarding screening factors and the hydrogenic values. In this paper we therefore put the GRASP2K-standard against two other recently proposed and implemented methods. The Welton interpretation~\cite{Welton1948} of the Self-energy which was implemented by Lowe et al.~\cite{Lowe2013} in the GRASP2K-package using the latest available hydrogenic values and modifying it to account for finite-nuclear-size effects. At about the same time
Shabaev et al.~\cite{Shabaev2013,Shabaev2015} developed a model QED approach to calculate the QED corrections to energy levels in relativistic many-electron atomic systems, which we have imported into the GRASP2K package.

Up to now, the spectral line from the $3d^9\ ^2D_{3/2}\rightarrow 3d^9\ ^2D_{5/2}$ transition has only been directly obsereved for seven of the Co-like ions(Zr$^{13+}$, Nb$^{14+}$, Mo$^{15+}$, Hf$^{45+}$, Ta$^{46+}$, W$^{47+}$ and Au$^{52+}$~\cite{Prior1987,Osin2012,Ralchenko2011}).

The aim of this work is to compare the importance of different contributions to the ground term fine structure in Co-like ions, as well as comparing different approaches to computing the Self-energy contribution. We will also discuss the possibility to distinguish between the results from different approaches with existing and possible future experimental measurements of this fine structure.

\section{\label{sec:calculation}calculation}
\subsection{\label{subsec:calculation}Correlation}
The MCDHF method implemented in the GRASP2K package starts from a Dirac-Coulomb Hamiltonian $H_{DC}$
\begin{equation}
H_{DC}= \sum_{i=1}^N(c~\bm{\alpha_i}\cdot\bm{p_i}+(\beta_i-1)c^2+V_i)+\sum_{i>i}^N\frac{1}{r_{ij}}
\end{equation}
where $V_i=-\frac{Z}{r_i}$ is the monopole part of the electron-nucleus interaction, $r_{ij}$ the distance between electrons $i$ and $j$, and $\alpha$ and $\beta$ are the Dirac matrices.
Electron correlation effect is included by expanding our Atomic State Function (ASF) $\Psi\left(\Gamma PJ\right)$ in a linear combination of Configuration State Functions (CSFs), $\Phi\left(\gamma_i PJ\right)$
\begin{equation}
\Psi\left(\Gamma PJ\right)=\sum_{i=1}^Mc_i\Phi\left(\gamma_i PJ\right)
\end{equation}
where $\gamma_i$ represents all other quantum numbers needed to uniquely define the CSF.
The CSFs are spin-angular coupled, antisymmetric products of Dirac orbitals
of the form
\begin{equation}
\phi({\bf r})=\frac{1}{r}\left(\begin{array}{c}P_{n\kappa}(r)\chi_{\kappa m                            }(\theta,\phi)\\
iQ_{n\kappa}(r)\chi_{-\kappa m}(\theta,\phi)\end{array}\right)
\end{equation}
The radial part of the one-electron orbitals and the expansion coefficients $c_i$ of the CSFs are obtained in the relativistic self-consistent field (RSCF) procedure. These are followed by a configuration interaction (RCI) approach, where Breit and QED effects are included. This implies that a limitation in the
GRASP2K package could be that the effects that are only included in the RCI step do not affect the orbitals, since they are not included in the RSCF procedure. To investigate the importance of this, we went back to the
results of a single-configuation approach (DHF), only including the $3d^9\ ^2D_{3/2}$ and $3d^9\ ^2D_{5/2}$ using the GRASP2K code and compared it to results from a B-spline version of a DHF program, DBSR-HF~\cite{Zatsarinny2016}, where the differential equations can be replaced by a set of generalized eigenvalue problems. In the latter code there are two options to include the Breit frequency-independent interaction, in the first one, 
it is added in the final stage as the GRASP2K code does, while in the second one,
it is included into orbital optimization.

\begin{figure}
  \includegraphics[width=\columnwidth]{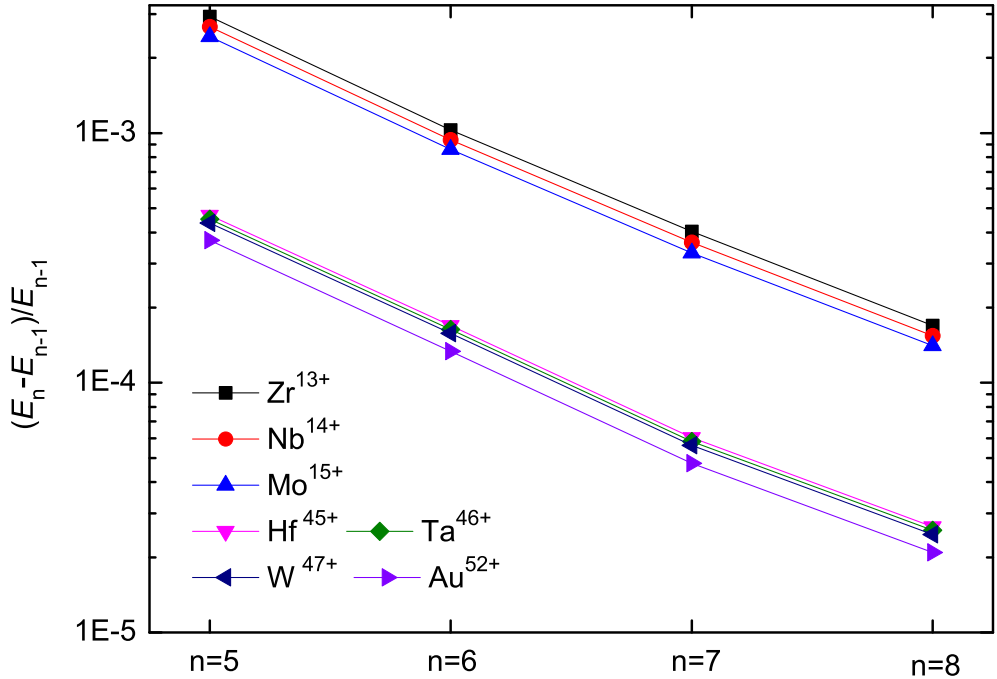}
  \caption{Convergence of the fine-structure energy as the function of largest $n$ in the active set.}\label{fig_converge}
\end{figure}

To represent correlation we start by using an extended version of the method from Guo et al.~\cite{Guo2016} to include valence and core-valence correlation (where the $3d$ is defined as the only valence subshell). In this we allow for single and double excitations from the $3d$ subshell, as well as single excitation from all core subshells ($3p, 3s, 2p, 2s, 1s$),
to an active set of orbitals with $l\le 5$ and $n\le 8$, to reach a clear convergence of electron correlation effect.
The convergence trend of the energy splittings for the ions of interest here is given in Fig.~\ref{fig_converge}. We can see that the energy splittings for the first 3 ions and the last 4 ions are converged to 0.015\% and 0.0025\%, respectively.
We can conclude, since the correlation is relatively small in the Co-like ions, that the "truncation" uncertainty in the computed fine structure, due to left out correlation, is negligible.

\subsection{\label{subsec:Breit}Breit Interaction}
The Transverse Photon interaction is included in GRASP2K through a standard Hamiltonian, as a correction to order $\alpha^2$
to the Dirac-Coulomb Hamiltionian included in the RSCF calculation
\begin{equation}
H_{Breit}=-\sum_{i<j}^{N}\left[\frac{\bm \alpha_i\cdot \bm \alpha_j}{r_{ij}}cos(\omega_{ij}r_{ij}/c)
+c^2(\bm\alpha_i\cdot\bm\bigtriangledown_i)(\bm\alpha_j\cdot\bm\bigtriangledown_j)
\frac{cos(\omega_{ij}r_{ij}/c)-1}{\omega_{ij}^2r_{ij}}\right]
\end{equation}
where $\omega_{ij}$ is the frequency of the exchanged virtual photon.
This reduces to the frequency-independent Breit interaction when $\omega_{ij}\rightarrow 0$ which we will label Breit(0).
The remaining and frequency-dependent part, we will label Breit($\omega$).
Breit(0) is the dominating correction to the Dirac-Coulomb results for all ions in the sequence and it is basically well understood how to include it in our calculations.
This is not the case for the Breit($\omega$) part, since the frequency is only representing a physical property for spectroscopic orbitals, i.e. the orbitals $1s, 2s, 2p, 3s, 3p$ and $3d$ that are occupied in the reference
term of $3d^9\ ^2D$. We therefore choose a method where the frequency-dependent contribution is only evaluated  between spectroscopic orbitals and put to zero if any others, often labeled correlation orbitals, are involved.

\subsection{\label{subsec:QED}QED Correction}
There are two contributions to the QED-correction, the Self-energy (SE) and the Vacuum Polarization (VP). For the Co-like ions, the SE dominates for all ions and we will therefore focus our investigation on different approaches to represent it. For the VP-contribtuion we evaluate the Uehling model potentials together with some higher order corrections~\cite{Fullerton1976}, as represented by the standard GRASP2K-approach.

In the current GRASP2K package, SE corrections are obtained based on a screened hydrogenic approximation
\begin{equation}
\Delta E_{SE}=\left(\frac{\alpha}{\pi}\right)\frac{\alpha^2Z^4}{n^3}F(nlj,Z\alpha)
\end{equation}
where $F(nlj,Z\alpha)$ is a slowly varying function of $Z\alpha$.
The total self-energy correction is given as a sum of one electron corrections weighted by the fractional occupation
number of the one-electron orbital in the wave function.
In the current GRASP2K package, the SE correction was included by relying on the hydrogenic results of Mohr et al.~\cite{Mohr1983} and Klarsfeld et al.~\cite{Klarsfeld1973}, the screening effect was included through a screened nuclear charge by taking the overlap integral of the wavefunction and a hydrogenic wavefunction. We will label the original GRASP2K calculation as `GRASP2K'.

By updating the GRASP2K program to use the latest available hydrogenic values~\cite{Mohr1992,LeBigot2001} and modifying it to account for finite-nuclear-size effects~\cite{Mohr1993,Beier1998}, Lowe et al.~\cite{Lowe2013} implemented a self-energy screening approximation based on the Welton interpretation~\cite{Welton1948}, it is labeled as `Welton' in the following.

Recently, Shabaev et al.~\cite{Shabaev2013,Shabaev2015} developed a model QED operator which also includes the non-local QED part to calculate the QED corrections for many-electron atomic systems.
Two nuclear model types are supported, the point nucleus and the extended nucleus, where we always use the latter.
We have included this model operator in the GRASP2K package and the results are labeled 'Shabaev'.

\section{\label{sec:results and discussion}results}
\subsection{\label{subsec:VBresults}Variational Breit}
Since a possible limitation of the GRASP2K package is the fact that Breit is not included in the variational RSCF calculation, and therefore do not affect the computed orbitals, but only affect the wave function through the mixing between different states, we compare in Table~\ref{tab_DBSR} the DHF results from the GRASP2K with our DBSR-HF calculations. In the latter the frequency-independent Breit interaction is either computed as in GRASP2K in a non-variation fashion - results labelled B(1) - or
included in the variational derivation of the orbitals in the method labelled B(2).
It is apparent from Table~\ref{tab_DBSR} that the DHF energy splittings calculated using the GRASP2K code and DBSR-HF B(1) code are in excellent agreement, as expected. The "variational effect" of the frequency-independent Breit, as manifested by the  differences between the B(1) and B(2) options of DBSR ranges from 1 cm$^{-1}$ for Zr$^{13+}$, to 67 cm$^{-1}$ for Au$^{52+}$. This is a rough or order of magnitude estimate of the uncertainty in our DHF calculations,
introduced by not including the Breit operator in the RSCF-part of our calculations. It is reasonable to
assume that the effect of this will decrease when correlation is included, since some of the effect will be incorporated in the expansion over CSFs.
As we will see this variational effect is even for DHF an order of magnitude smaller than contributions from correlation and will therefore not affect the conclusions of this paper.

\begin{table}
\small
\caption{\label{tab_DBSR}%
The DHF energy splittings and the contributions of frequency-independent Breit interaction calculated using the GRASP2K code and DBSR-HF code.}
\begin{ruledtabular}
\begin{tabular}{ccrrrrrrr}
  &  &Zr$^{13+}$ & Nb$^{14+}$ & Mo$^{15+}$& Hf$^{45+}$&Ta$^{46+}$& W$^{47+}$ & Au$^{52+}$\\
\colrule
GRASP2K&DHF  & 20841.3 & 24166.0 & 27863.0 & 479335.9 & 512732.8 & 547862.1 & 751681.5 \\
&Breit(0) & -869.3  & -973.9  & -1086.6 & -10161.4 & -10731.1 & -11324.0 & -14658.1 \\
DBSR-HF &DHF  & 20841.2 & 24165.8 & 27862.9 & 479334.6 & 512731.4 & 547860.6 & 751679.3 \\
&B(1)   & -869.3  & -973.9  & -1086.6 & -10161.4 & -10731.1 & -11324.0 & -14658.1 \\
&B(2)   & -868.3  & -972.7  & -1085.1 & -10119.8 & -10686.4 & -11276.1 & -14591.1 \\
&$\delta_{var}$& 1.0  &   1.2   & 1.5 &        41.6  &     44.7 &    47.9  &  67.0 \\
\end{tabular}
\end{ruledtabular}
\end{table}
\subsection{The Self-energy contribution.}\label{SE}
\begin{figure}
  \includegraphics[width=0.7\columnwidth]{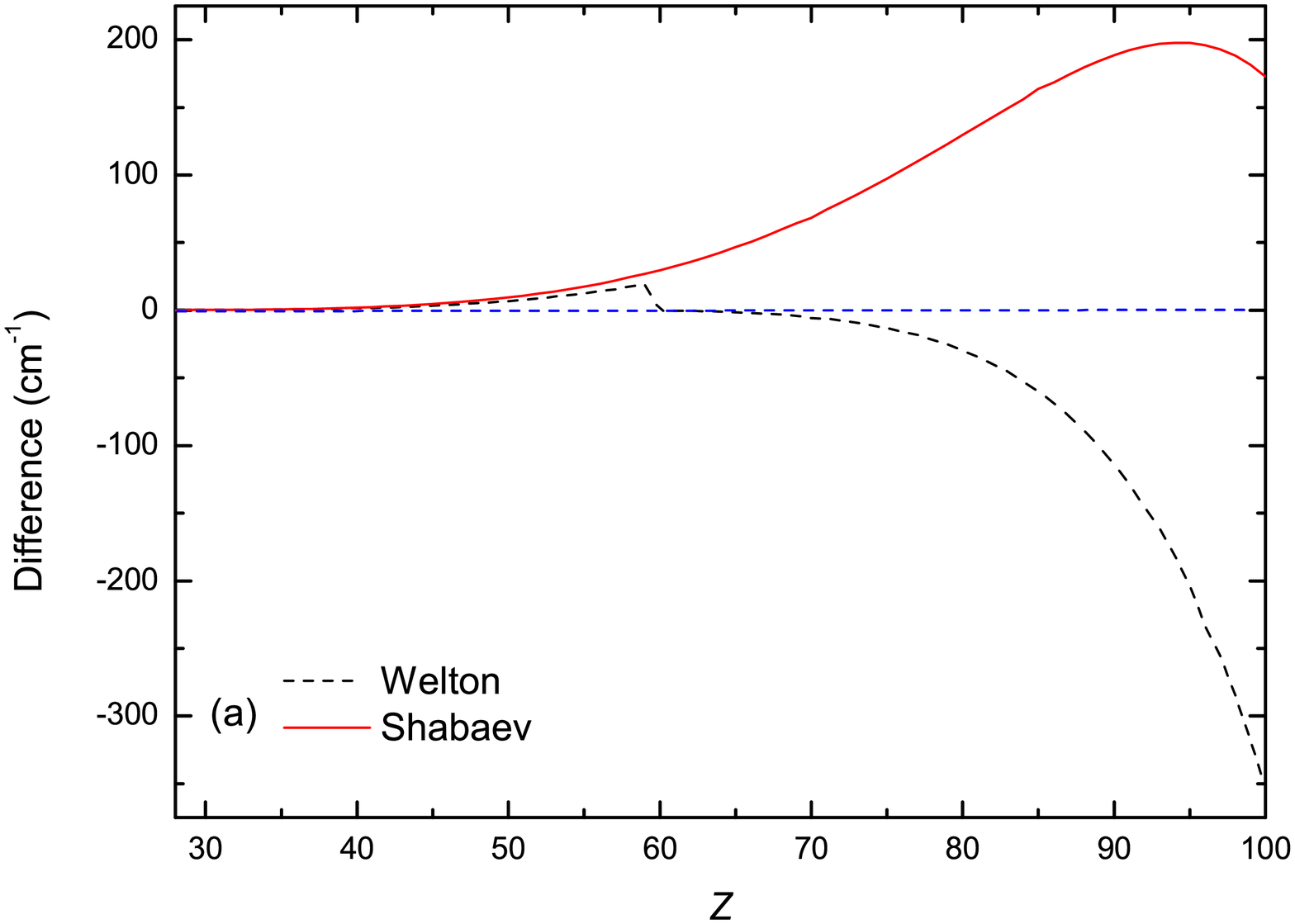}

\includegraphics[width=0.7\columnwidth]{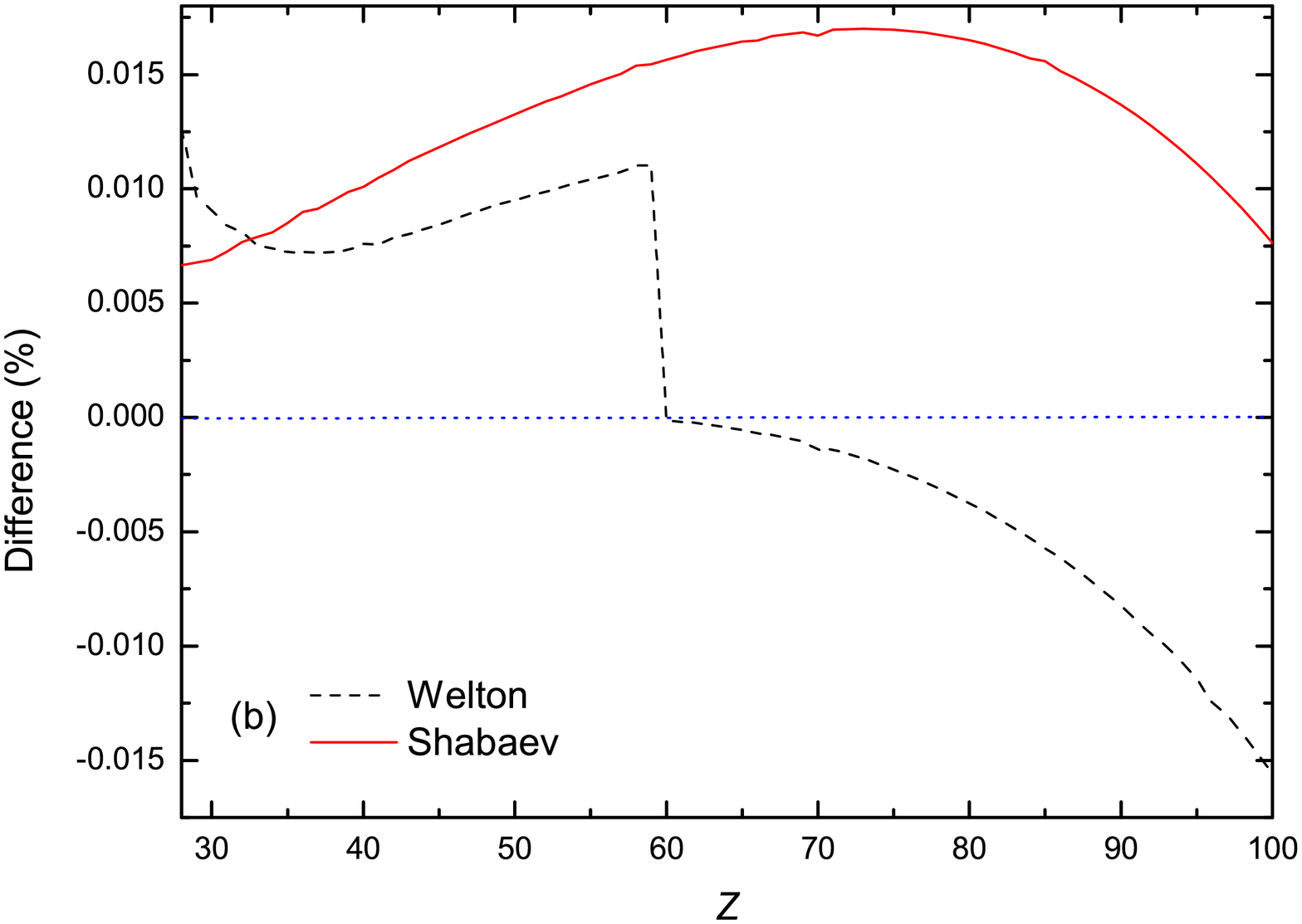}
  \caption{Differences ((a) in cm$^{-1}$, (b) in \%) in contributions of SE to the energy splittings for Co-like ions from calculations based on Welton's and Shabaev's concepts compared to the original GRASP2K result. For en explanation of the jump in the curve for Welton's method, see section~\ref{SE}.  The horizontal lines at 0 represents th GRASP2K results.
}\label{fig_Co-SE}
\end{figure}

As discussed above, we use three different approximations to deal with the SE corrections, based on standard GRASP2K~\cite{Jonsson2013}, the Welton interpretation~\cite{Welton1948,Lowe2013} and the Shabaev approach~\cite{Shabaev2013,Shabaev2015}. These are compared in Fig.~\ref{fig_Co-SE}, and it is clear that the results based on the Shabaev-method are larger than the GRASP2K results for all ions, and their differences smoothly vary with increasing atomic number.
The SE contributions calculated using Welton's concept are also larger than the GRASP2K results for $Z\leq60$ but then decreases to become smaller for larger $Z$. There is a significant jump for the differences between these two calculations at $Z=60$, which is due to the fact that the SE contributions from electrons with $n=3, 4, 5$ and $5/2\leq j\leq 9/2$ for $60\leq Z\leq110$ were taken from the results of Le Bigot et al.~\cite{LeBigot2001}, but those for $Z<60$ were obtained from an extrapolation.

\subsection{Contributions to the fine structure.}
\begin{figure}
  \includegraphics[width=0.85\columnwidth]{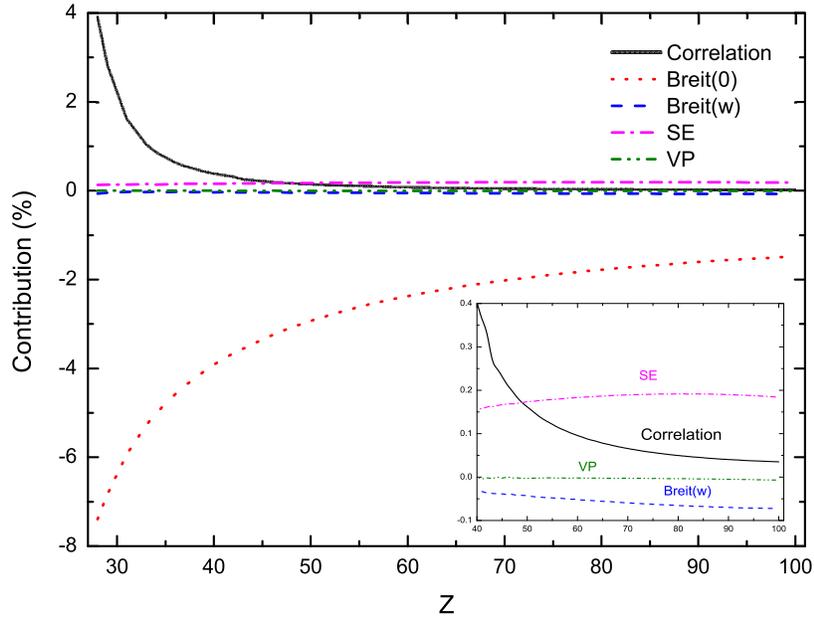}
  \caption{Contributions from different effects to the fine structure splitting of the ground term in Co-like ion, where Breit(0) represents frequency-independent Breit, Breit($\omega$) frequency-dependent Breit, SE self-energy and VP vacuum polarization.}\label{fig_Co}
\end{figure}

Table~\ref{tab_Co-like} and Figure~\ref{fig_Co} present different contributions to the energy splitting for a number of ions, where direct measurements of the fine structure is available.
It is clear that Breit(0) represents the largest contribution, followed by Self-energy, for most ions.
Since the correlation and Breit($\omega$) have opposite signs for large Z, and almost cancel each other, these ions are excellent testing ground for the remaining Self-energy contribution.

\begin{table}
\small
\caption{\label{tab_Co-like}%
Contributions to the fine structure splittings of the $3d\ ^2D$ term. Breit(0) and Breit($\omega$) represents frequency-independent and -dependent Breit-interaction, respectively. Self-energy contributions are given for the original GRASP2K method~\cite{Jonsson2013} (GRASP2K), according to Welton's concept~\cite{Welton1948,Lowe2013} (WELTON) and the model operator approach~\cite{Shabaev2013,Shabaev2015} (SHABAEV). The experimental error estimates ($\delta_{exp}$) are from references given in Table~\ref{tab_Co-like_tot}. All results are given in cm$^{-1}$.}
\begin{ruledtabular}
\begin{tabular}{crrrrrrr}
   &Zr$^{13+}$& Nb$^{14+}$& Mo$^{15+}$&Hf$^{45+}$&Ta$^{46+}$&W$^{47+}$&Au$^{52+}$\\
\colrule
Correlation    & 79.9   & 83.8    & 87.9    & 321.0    & 333.2    & 345.6    & 413.6   \\
Breit(0)       & -816.0 & -914.0  & -1019.4 & -9398.1  & -9919.6  & -10462.0 & -13505.5 \\
Breit($\omega$)& -7.2   & -8.5    & -10.3   & -289.6   & -313.0   & -337.8   & -488.2  \\
SE\_GRASP2K    & 30.9   & 36.2    & 42.1    & 831.3    & 890.8    & 953.4    & 1317.2   \\
SE\_WELTON    & 32.3   & 37.9    & 44.2    & 823.8    & 881.7    & 942.5    & 1292.6  \\
SE\_SHABAEV    & 32.9   & 38.6    & 45.0    & 911.2    & 976.4    & 1044.8   & 1440.2 \\
$\delta_{exp}$& 1     & 5       & 2       & 67       & 76       & 87       & 164     \\
\end{tabular}
\end{ruledtabular}
\end{table}
\subsection{Final results and comparison with experiments.}
\begin{figure}
  \includegraphics[width=\columnwidth]{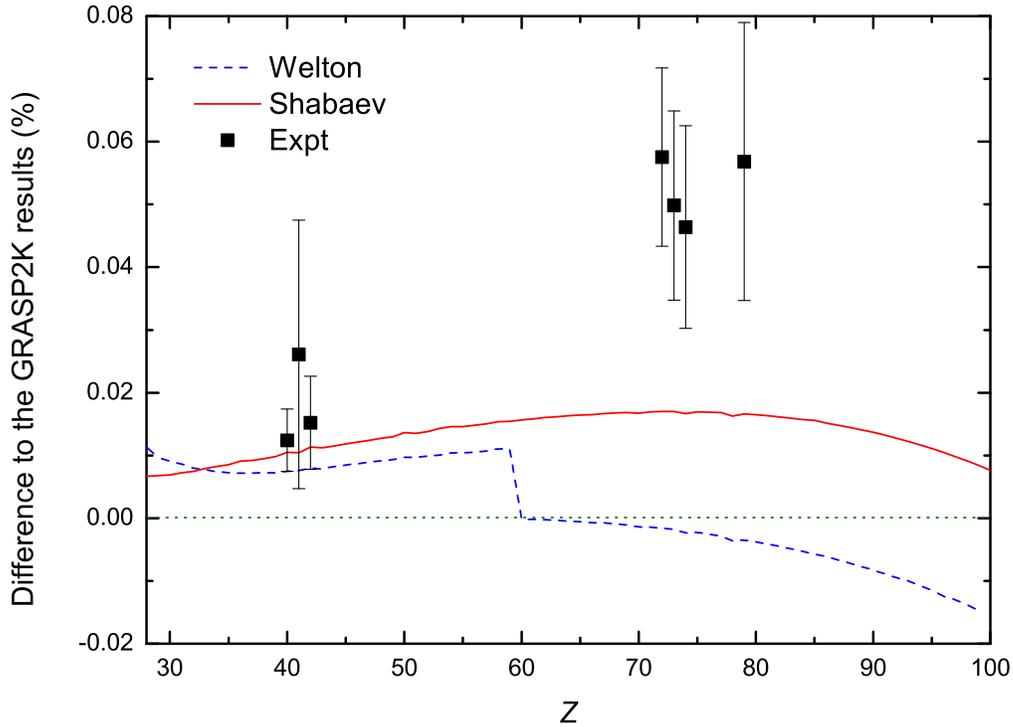}
  \caption{Comparison of the fine structure energy
splittings from direct observations (see text) with
the results of the present calculations. The horizontal lines at 0 represents the GRASP2K results.}\label{fig_Co_tot}
\end{figure}

In Table~\ref{tab_Co-like_tot} and Figure~\ref{fig_Co_tot} we present the final results for our three different models for Self-energy, compared with experimental observations.
For the first three ions, the final results for all models are all within or close to the experimental error bars. However, for the four high-$Z$ ions, none of the three calculations agree with experiment to within the experimental error bar, albeit the Shabaev-method is closest, being lower than the experimental values by 0.03\%$-$0.04\%.
It is clear that this can not be attributed to correlation, since for these four ions, the convergence of our present calculation is about 25 ppm and the left-out core-core (CC) correlation has en estimated, maximal contribution of only 30 ppm~\cite{Guo2016a}.

The final results for the whole isoelectronic squence with $28\leq Z\leq100$  are listed in Table~\ref{tab_full}. Although except for the 7 ions where direct measurements of the fine structure is available, the active set for $28\leq Z \leq42$ is restricted to $n\leq 7, l\leq5$, while for $43\leq Z \leq100$ is $n\leq 6, l\leq5$, as shown in figure~\ref{fig_converge} and our previous paper~\cite{Guo2016}, the convergence of the electron correlation calculations is within 0.01\%.

\begin{table}
\small
\caption{\label{tab_Co-like_tot}%
The calculated energy splittings with different treatments of Self-energy (see Table~\ref{tab_Co-like} for explanations of notations) compared with the experimental values from direct observations. The experimental values are from Ref.\cite{Suckewer1982} for Zr$^{13+}$ and Mo$^{15+}$; Ref.~\cite{Prior1987} for Nb$^{14+}$; Ref.~\cite{Osin2012} for Hf$^{45+}$, Ta$^{46+}$ and Au$^{52+}$; Ref.~\cite{Ralchenko2011} for W$^{47+}$. All results are in cm$^{-1}$.}
\begin{ruledtabular}
\begin{tabular}{crrrrrrr}
   &Zr$^{13+}$& Nb$^{14+}$& Mo$^{15+}$&Hf$^{45+}$&Ta$^{46+}$&W$^{47+}$&Au$^{52+}$\\
\colrule
GRASP2K        & 20128.5& 23362.9 & 26962.9 & 470783.1 & 503705.0 & 538340.3 & 739388.8\\
WELTON         & 20130.0& 23364.6 & 26965.0 & 470775.6 & 503695.9 & 538329.4 & 739363.4\\
SHABAEV        & 20130.5& 23365.3 & 26965.8 & 470863.0 & 503790.6 & 538431.7 & 739511.8\\
EXPT            & 20131(1)  & 23369(5)   & 26967(2)   & 471054(67)   & 503956(76)   & 538590(87)   & 739810(164)  \\
\end{tabular}
\end{ruledtabular}
\end{table}

\newpage
\begin{longtable}{rrrrrrrr}
\caption{Calculated $3d^9\ ^2D_{5/2,3/2}$ fine-structure splittings (in cm$^{?1}$) for Co-like ions with 28$\leq Z \leq$100. (see Table~\ref{tab_Co-like} for explanations of notations). Results are in cm$^{-1}$.}
\label{tab_full}\\
\hline
 $Z$  & GRASP2K & WELTON & SHABAEV & $Z$  & GRASP2K & WELTON & SHABAEV \\
\hline
\endfirsthead
\multicolumn{6}{l}{Table~\ref{tab_full}, (continued)}\\
\hline
 $Z$  & GRASP2K & WELTON & SHABAEV & $Z$  & GRASP2K & WELTON & SHABAEV \\
\hline
\endhead
\hline
\endfoot
28	&	1503.7	&	1503.8	&	1503.8	&	65	&	283426.6	&	283425.1	&	283473.3	\\
29	&	2068.5	&	2068.7	&	2068.7	&	66	&	305948.6	&	305946.4	&	305999.0	\\
30	&	2756.8	&	2757.1	&	2757.0	&	67	&	329792.8	&	329790.3	&	329847.8	\\
31	&	3581.5	&	3581.9	&	3581.8	&	68	&	355012.3	&	355009.1	&	355071.8	\\
32	&	4558.6	&	4559.0	&	4559.0	&	69	&	381660.3	&	381656.3	&	381724.6	\\
33	&	5705.5	&	5705.9	&	5706.0	&	70	&	409792.3	&	409786.5	&	409860.7	\\
34	&	7040.1	&	7040.7	&	7040.7	&	71	&	439461.4	&	439455.2	&	439535.9	\\
35	&	8581.7	&	8582.4	&	8582.5	&	72	&	470783.1	&	470775.6	&	470863.0	\\
36	&	10350.1	&	10350.9	&	10351.1	&	73	&	503705.0	&	503695.9	&	503790.6	\\
37	&	12366.6	&	12367.5	&	12367.7	&	74	&	538340.3	&	538329.4	&	538431.7	\\
38	&	14652.5	&	14653.5	&	14653.9	&	75	&	574688.0	&	574675.0	&	574785.5	\\
39	&	17230.6	&	17231.8	&	17232.3	&	76	&	612931.7	&	612916.2	&	613035.3	\\
40	&	20128.5	&	20130.0	&	20130.5	&	77	&	653074.9	&	653056.5	&	653184.8	\\
41	&	23362.9	&	23364.6	&	23365.3	&	78	&	695184.9	&	695160.1	&	695298.2	\\
42	&	26962.9	&	26965.0	&	26965.8	&	79	&	739388.8	&	739363.4	&	739511.8	\\
43	&	30937.5	&	30940.0	&	30941.0	&	80	&	785551.5	&	785521.9	&	785681.1	\\
44	&	35347.9	&	35350.8	&	35352.0	&	81	&	833950.4	&	833916.0	&	834086.6	\\
45	&	40205.7	&	40209.1	&	40210.4	&	82	&	884589.0	&	884544.1	&	884726.7	\\
46	&	45539.8	&	45543.8	&	45545.5	&	83	&	937523.3	&	937477.6	&	937672.8	\\
47	&	51381.1	&	51385.6	&	51387.4	&	84	&	992841.1	&	992788.6	&	992997.0	\\
48	&	57759.8	&	57765.1	&	57767.2	&	85	&	1050611.4	&	1050551.4	&	1050775.0	\\
49	&	64708.4	&	64714.4	&	64716.8	&	86	&	1110909.3	&	1110841.0	&	1111077.6	\\
50	&	72259.5	&	72266.5	&	72269.4	&	87	&	1173814.0	&	1173736.1	&	1173988.0	\\
51	&	80448.0	&	80455.8	&	80458.9	&	88	&	1239399.4	&	1239310.8	&	1239578.8	\\
52	&	89308.1	&	89316.9	&	89320.4	&	89	&	1307746.3	&	1307645.8	&	1307930.7	\\
53	&	98876.0	&	98886.0	&	98890.2	&	90	&	1378937.3	&	1378823.4	&	1379126.0	\\
54	&	109188.6	&	109200.0	&	109204.5	&	91	&	1453052.0	&	1452923.3	&	1453244.3	\\
55	&	120284.7	&	120297.2	&	120302.2	&	92	&	1530178.5	&	1530033.3	&	1530373.6	\\
56	&	132202.6	&	132216.5	&	132222.1	&	93	&	1610396.9	&	1610235.5	&	1610593.9	\\
57	&	144982.6	&	144998.1	&	145004.4	&	94	&	1693800.0	&	1693618.5	&	1693997.9	\\
58	&	158665.7	&	158683.2	&	158690.1	&	95	&	1780469.7	&	1780265.9	&	1780667.2	\\
59	&	173294.8	&	173313.9	&	173321.6	&	96	&	1870500.7	&	1870267.5	&	1870696.7	\\
60	&	188912.6	&	188912.4	&	188942.2	&	97	&	1963982.1	&	1963726.6	&	1964175.0	\\
61	&	205563.7	&	205563.3	&	205596.2	&	98	&	2061009.1	&	2060723.8	&	2061197.3	\\
62	&	223293.4	&	223292.9	&	223329.2	&	99	&	2161672.7	&	2161354.6	&	2161854.2	\\
63	&	242148.6	&	242147.8	&	242187.8	&	100	&	2266073.7	&	2265719.7	&	2266246.4	\\
64	&	262176.8	&	262175.6	&	262219.5	&		&		&		&		\\
\hline
\end{longtable}


\section{\label{sec:conclusion}conclusion}
In this paper, using the Co-like ions where the correlation effect is small as an example, we investigate the contributions from different physical effects to the ground configuration's fine structure splittings.
It shows that the frequency independent Breit contribution Breit(0) has the largest contribution for all ions, correlation effect decreases fast with $Z$, the frequency dependent Breit contribution Breit($\omega$) is non-negligible especially for high-$Z$ ions and self energy becomes the largest correction for $Z>50$.

We estimated the SE correction using three approximation methods and showed that the Model QED operator results labeled Shabaev shows the best agreement with experimental values, but is outside the error bars for high-$Z$. We argue that this is not due to errors in the correlation treatment or from excluding the Breit operator in the variation RSCF-procedure. The reason could possibly be found in the treatment of the frequency-dependent Breit-interaction, but more likely in the treatment of the Self-energy, which dominates for these ions. Another possibility would be larger uncertainties than stated in the experiments.

It is clear that these systems could be used to probe the method for computing Breit and QED corrections and accurate experiments could distinguish between different computational approaches.

We recommend that these systems are revisited both experimentally, with larger accuracy in the direct measurements of the fine structure, and in modeling of Breit and QED-effects.

\begin{acknowledgments}
We acknowledge the support from the National Natural Science Foundation of China Grant No. 11674066 and Grant No. 11474069, the Swedish Research Council (VR) under Contract No. 2015-04842 and the Canada NSERC Discovery Grant 2017-03851.
\end{acknowledgments}

\clearpage
\bibliographystyle{aip}
\bibliography{Co-like}

\end{document}